\documentclass{biostatistics}
\usepackage[hyphens]{url}
\usepackage{graphicx}
\RequirePackage{amsthm,amsmath,amsfonts,amssymb}
\usepackage{natbib}
\RequirePackage{bm}
\RequirePackage{lastpage}
\usepackage{multirow}
\usepackage{multicol}
\usepackage{enumerate}
\usepackage{cprotect}
\usepackage[algosection,linesnumbered,lined,boxed,commentsnumbered]{algorithm2e}
\usepackage{float}
\usepackage[caption=false]{subfig}
\usepackage{booktabs}
\usepackage[table,xcdraw]{xcolor}

\usepackage{xr}
\usepackage{xcite}
\externalcitedocument{supplement}

\def\mcL{\mathcal{L}}
\def\mcT{\mathcal{T}}

\newcommand{\Real}{\mathbb{R}}

\newcommand{\tr}{\operatorname{tr}} 

\def\bG{\mathbf{G}}

\def\bX{\mathbf{X}}
\def\bY{\mathbf{Y}}
\def\bE{\mathbf{E}}

\def\bB{\mathbf{B}}

\def\bG{\mathbf{G}}

\def\bX{\mathbf{X}}
\def\bY{\mathbf{Y}}
\def\bE{\mathbf{E}}

\def\bB{\mathbf{B}}

\def\bphi{\boldsymbol{\phi}}
\def\bpsi{\boldsymbol{\psi}}
\def\bxi{\boldsymbol{\xi}}
\def\bmu{\boldsymbol{\mu}}

\def\bGamma{\boldsymbol{\Gamma}}

\def\bSigma{\boldsymbol{\Sigma}}

\theoremstyle{plain}

\newtheorem{prop}[theorem]{Proposition}

\newcommand{\KL}{Karhunen-Lo\'eve }

\begin{document}

\title{Projection-based two-sample inference for sparsely observed multivariate functional data}

\author{Salil Koner$^\ast$ and Sheng Luo\\[6pt]
\textit{Department of Biostatistics and Bioinformatics}\\
\textit{Duke University, Durham, NC, USA}
\\[2pt]
{salil.koner@duke.edu, sheng.luo@duke.edu}}

\markboth%
{Koner and Luo}
{Two-sample inference for sparsely observed multivariate functional data}

\maketitle

\footnotetext{Salil Koner, email: salil.koner@duke.edu}

\begin{abstract}
{Modern longitudinal studies collect multiple outcomes as the primary endpoints to understand the complex dynamics of the diseases. Oftentimes, especially in clinical trials, the joint variations among the multidimensional responses play a significant role in assessing the differential characteristics between two or more groups, rather than drawing inferences based on a single outcome. Enclosing the longitudinal design under the umbrella of sparsely observed functional data, we develop a projection-based two-sample significance test to identify the difference between the typical multivariate profiles. The methodology is built upon widely adopted multivariate functional principal component analysis to reduce the dimension of the infinite-dimensional multi-modal functions while preserving the dynamic correlation between the components. The test is applicable to a wide class of (non-stationary) covariance structures of the response, and it detects a significant group difference based on a single p-value, thereby overcoming the issue of adjusting for multiple p-values that arises due to comparing the means in each of components separately. Finite-sample numerical studies demonstrate that the test maintains the type-I error, and is powerful to detect significant group differences, compared to the state-of-the-art testing procedures. The test is carried out on the longitudinally designed TOMMORROW study of individuals at high risk of mild cognitive impairment due to Alzheimer's disease to detect differences in the cognitive test scores between the pioglitazone and the placebo groups.}
{Alzheimer's disease; 
Multivariate longitudinal data; TOMMORROW trial, Sparse functional data;}
\end{abstract}

\section{Introduction} \label{sec: mface_testing_intro}

Two-sample inference problem is ubiquitous in many biostatistical applications such as drug discovery, and assessing the target drug's efficacy, as compared to placebo or a standard of care treatment. To characterize Alzheimer's disease (AD), a complex chronic disease, multivariate longitudinal data are always collected. In the setting of randomized controlled trials (RCT), it is necessary to test the treatment efficacy across multiple longitudinal outcomes. For example, in the TOMMORROW study, several cognitive testing battery scores are collected to evaluate the efficacy of pioglitazone to delay mild cognitive impairment due to Alzheimer's disease (AD) \citep{burns2019tommorrow, burns2021safety}. Traditional methodologies summarize these multiple subject-level measurements to obtain a single composite score, and conduct the two-sample inference problem based on the derived score. Such summary-based approaches completely overlook the dynamic variation among the multidimensional endpoints collected for each subjects, thereby fail to quantify how the influence of the study drug on one endpoint affects the other. Several of these approaches further simplify the inference problem by focusing only on the change of the outcome between baseline and the last visit of the study, thus completely ignoring the longitudinal nature of the outcome \citep{kieburtz2015effect}. Therefore, inference procedures that simultaneously account for this joint variation among multiple endpoints associated to disease progression as well as the longitudinal nature of the outcomes is crucial for powerful detection of significant group differences.

This article focuses on two-sample inference problem for repeatedly measured multidimensional outcome. In the context of the motivating TOMMORROW study with a cohort of high-risk at Mild Cognitive Impairment (MCI) subjects, twelve questionnaire based self-reported cognitive performance scores are collected repeatedly over time at the individual level. These twelve scores correspond to five major cognitive domains, namely, episodic memory, perceptual memory, semantic memory, working memory, and visuospatial memory. The subjects in the study are given daily oral $0.8$ mg of pioglitazone or placebo. The goal is to assess the effectiveness of pioglitazone in delaying the onset of MCI due to AD in high-risk cognitive-normal subjects, as indicated by the longitudinally measured 12-tuple cognitive scores corresponding to the five primary cognitive domains. The duration of the study is five years, and the individuals are expected to follow up every six months. Further details on the design of the study are reported in Section~\ref{sec: mface_testing_azillectstudy} and \cite{burns2021safety}.  Prevalent to modern longitudinal studies, the individuals fill up the questionnaires at their own time; as a result, the specific time points (as measured from the baseline) at which the responses are reported can be different for each subject.

We approach the two-sample inference problem within a framework of multivariate functional data (mv-fD), where the response is the trajectory of the multidimensional cognitive scores observed for a subject over the span of five years. Contrary to the traditional mixed model framework that quantifies the temporal dependence by imposing some structural assumptions such as linear or quadratic in time, functional data analysis (fDA) methods relax these structural conditions by assuming smoothness in the 
mean and covariance of the response trajectories.
However, existing inference procedures for mv-fD primarily concentrate on the setting when the functions are observed densely, i.e., when the trajectories are fully observed. \cite{gorecki2017multivariate} first approached this problem by expanding each individual functional trajectory into a set of basis functions, and constructed several multivariate analysis of variance (MANOVA) type test-statistics based on the random coefficient of the basis representations. \cite{qiu2021two} considered a pointwise Hotelling $T$-squared statistic to test the null hypothesis of no group difference. \cite{jiang2017two} considered a energy-based distance between the empirical characteristic functions of the samples in the two groups. \cite{krzysko2021two} constructed a test of conformity by considering the divergence between the characteristic functions of the coefficients of the basis expansions. A robust analysis of variance for two-sample inference on mv-fD using multivariate statistical depth function \citep{claeskens2014multivariate} is developed by \cite{qu2021robust}. These procedures require the individual functions to be observed fully under a dense or ultra-dense design, and can not be applied to the longitudinally designed TOMMORROW study, where the outcomes are recorded sparsely for each subject, with only a few number of recordings may be available for some subjects.

The article proposes a two-sample testing procedure for {\em sparsely observed multivariate functional data}. Our method relies upon projecting the multidimensional functions jointly onto a set of data-driven multivariate orthogonal bases, and use the projections to test for the group difference. The projections are obtained parsimoniously by representing the data in the form of multivariate \KL expansion, along the principal directions of variation. Using these projections as a proxy to the entire multidimensional data, we compare them to test for a significant group difference as a whole based on one single p-value. Thus, our projection-based approach is suitable to detect a significant group difference whenever the two groups are different from each other with respect to any of the component functions (i.e., any of the 5 cognitive scores as in the TOMMORROW study). In contrast, separate component-wise tests for group difference would lead to handling multiple p-values to adjust for the global type I error rate \citep{pomann2016two, wang2021two}. We further present the asymptotic null distribution of the test, as well as provide a permutation-based mechanism for approximating the null distribution in finite sample scenario. The numerical studies demonstrate that the proposed test maintains the type-I error, and is powerful to detect a slight departure from the null hypothesis compared to the alternative. The procedure is computationally fast as a function of sample size, is also scalable to the dimension of the multivariate outcomes. 

The rest of the paper is organized as follows. In section~\ref{sec: mface_testing_Statframework} we introduce the framework of multivariate functional data, and mathematically formulate the hypotheses. We develop the projection-based testing framework in Section~\ref{sec: projectionbasedinference}, and present the test statistic along with the asymptotic null distribution in Section~\ref{sec: HTteststatistic}. Numerical studies in Section~\ref{sec: mface_testing_simstudy} demonstrates the empirical size and power properties of the test to detect significant group difference. Section~\ref{sec: mface_testing_azillectstudy} discusses the findings obtained from the application of our test on TOMMORROW study.

\section{Statistical Framework} \label{sec: mface_testing_Statframework}


For each participant $i=1,\dots, n$, let $Y_{ij}^{(\ell)}$ denote the $\ell$th response measured at time $t_{ij}$ from the baseline, $\ell=1,\dots,q$, associated to $j$th clinic visit, $j=1,\dots, m_i$. So, there are $n$ subjects, and a $q$-dimensional response vector is recorded for subject $i$ at each of $m_i$ observation points. In our data application, we model the scores corresponding to five cognitive domains, namely episodic, perceptual, semantic, working, and visuospatial as $Y_{ij}^{(\ell)}$, $\ell=1, \dots, 5$ respectively. Denote by the bold symbol ${\bY}_{ij}:= (Y_{ij}^{(1)}, \dots, Y_{ij}^{(q)})^\top$, a collection of $q$-dimensional response jointly recorded at time $t_{ij}$ for $i$th subject. Throughout the article, we will assume that $q$ is fixed, and does not grow with $n$. The data structure is unbalanced, i.e. the observation points $\{t_{i1}, \dots, t_{im_i}\}$ at which the measurements are recorded for $i$th subject, and the number of repeated measures, $m_i$, are different for each subject.  For $\ell=1, \dots, q$, we model $\{Y_{ij}^{(\ell)}: j=1, \dots, m_i\}$ as noisy realization of a smooth latent process $X_i^{(\ell)}(\cdot)$ defined on a compact domain $\mcT \subset \Real$, evaluated at $\{t_{i1}, \dots, t_{im_i}\}$, i.e.
\begin{equation} \label{eqn: model}
    {Y}^{(\ell)}_{ij} = X_i^{(\ell)}(t_{ij}) + \epsilon_{ij}^{(\ell)},
\end{equation}
for some measurement error $\epsilon_{ij}^{(\ell)}$ that are independently distributed across $i$, $j$, and $\ell$. We assume that $\epsilon_{ij}^{(\ell)} \sim (0, \tau_\ell^2)$, $\ell=1, \dots, q$. For each participant $i$, we assume that the latent process $X_i(\cdot)$ is sparsely observed, i.e. $m_i$ is small. Denoting by bold symbol $\bX_i(\cdot) :=(X_i^{(1)}(\cdot), \dots, X_i^{(q)}(\cdot)) $, the vector of $q$-dimensional latent process, we assume that for each $i$,
$\bX_i(\cdot)$ belongs to $\mcL^2[\mcT]$ - the space of all square-integrable $q$-dimensional random functions in $\mcT$, associated with the inner product $\langle \mathbf{f}_1, \mathbf{f}_2 \rangle = \sum_{\ell=1}^q \langle f_1^{(\ell)} , f_2^{(\ell)}\rangle = \sum_{\ell=1}^q \int_{\mathcal{T}} f_1^{(\ell)}(t)f_2^{(\ell)}(t)dt$, for $\mathbf{f}_1$ and $\mathbf{f}_2$ in $\mcL^2(\mcT)$.  

Let $g_i$ be the group indicator of the $i$th subject, i.e. $g_i=1$ if subject $i$ is in treatment (pioglitazone) group, and $g_i = 0$ for the placebo group. Denote the mean of $\bX_i(\cdot)$ by
\begin{equation} \label{eqn: meanofX}
    \mathbb{E}[\bX_i(t)] = \boldsymbol\mu(t) + g_i \boldsymbol\eta(t),
\end{equation}
where $\mathbb{E}\,[X_i^{(\ell)}(t)] = \mu^{(\ell)}(t) + g_i \eta^{(\ell)}(t)$, $\ell=1,\dots,q$ be the outcome-specific mean. Therefore, $\mu^{(\ell)}(t)$ represents general mean response for $\ell$th response at time $t$, and the $\eta^{(\ell)}(t)$ represents the difference in the mean for $\ell$th response between the treatment and placebo group at time $t$. Furthermore, define the covariance operator of process by $\boldsymbol\Xi$, where for any $\mathbf{f} \in \mcL^2(\mcT)$, the operator $(\boldsymbol\Xi \,\mathbf{f})(t)$ has elements $(\boldsymbol\Xi\,\mathbf{f})^{(\ell)}(t) = \sum_{\ell^\prime=1}^q \langle \Sigma_{\ell\ell^\prime}(t, \cdot)\;,f^{(\ell^\prime)} \rangle$, with $\Sigma_{\ell\ell^\prime}(t, t^\prime) = \mathrm{Cov}\{X_i^{(\ell)}(t), X_i^{(\ell^\prime)}(t^\prime)\}$ for $1 \leq \ell,\ell^\prime \leq q$ being the covariance kernel between $\ell$ and $\ell^\prime$th component of $\bX_i(t)$. For fixed $t, t^\prime \in \mcT$, the $q \times q$ covariance matrix between $\bX_i(t)$ and $\bX_i(t^\prime)$ can be expressed as $ \text{Cov}({\bX}_i(t), {\bX}_i(t^\prime)) = {\boldsymbol\Sigma}(t,t^\prime) = \{\Sigma_{\ell\ell^\prime}(t,t^\prime)\}_{1 \leq \ell, \ell^\prime \leq q}$. Under the above notations, the null hypothesis of testing the significance of the treatment effect mathematically converts to,
\begin{align*}
 &H_0: \boldsymbol\eta(t) = (\eta^{(1)}(t), \dots, \eta^{(q)}(t))^\top  = 0 \quad \forall \;\; t \in \mcT, \\
 & \hspace{1.5 in } \textrm{versus} \\
 &H_1: \eta^{(\ell)}(t) \neq 0 \quad \text{for some } t \in \mathcal{T}, \text{ and for some } \ell =1, \dots, q.
\end{align*}

  A primitive way to test the null hypothesis involving a $q$-dimensional functional effect is to test component-wise, i.e. to test whether each of $\eta^{(\ell)}(t) = 0 \; \forall \; t$, for all $\ell=1, \dots, q$. Several methods for testing significance of univariate functional effect have been proposed in the literature, such as the divergence-based statistic of \cite{zhang2007statistical}, intended for densely observed functions; likelihood ratio based tests have been proposed by \cite{crainiceanu2004likelihood} to accommodate for sparse functional data. However, these separate component-wise tests do not account for the joint dynamic variation among the multivariate components. Moreover, there is no general way to pool the individualized inferences drawn from each tests to further conclude about the global null hypothesis, especially when the number of longitudinal outcomes is large. We address these issues in our projection-based multivariate joint testing procedure described in Section~\ref{sec: projectionbasedinference}.
 


\section{Multivariate projection-based inference approach} \label{sec: projectionbasedinference}

Let $\{\boldsymbol{\phi}_1(\cdot), \dots,\boldsymbol{\phi}_k(\cdot), \dots \}$ be a set of orthonormal basis function for $\mcL^2[\mcT]$ with $\langle \bphi_k, \bphi_{k'} \rangle =\mathbb{I}(k=k')$ for $k, k'\geq 1$. Here $\mathbb{I}(\cdot)$ stands for the indicator function. The treatment effect $\boldsymbol\eta(t)$ can be represented uniquely as $ \boldsymbol\eta(t) = \sum_{k=1}^{\infty} c_k\bphi_k(t)$ for all $t \in \mathcal{T} $, where $c_k = \langle \boldsymbol\eta, \bphi_k\rangle = \sum_{\ell=1}^q\langle \eta^{(\ell)}, \phi_k^{(\ell)}\rangle $ is the projection of $\boldsymbol\eta(t)$ onto $\bphi_k(t)$, for $k\geq 1$. This implies that the square of the functional norm of $\boldsymbol\eta$ is $\lVert \boldsymbol\eta \rVert^2 = \sum_{k=1}^\infty c_k^2$. Note that, under the null hypothesis, $c_k = 0$ for all $k \geq 1$. On the other hand, under $H_1$, there exists at least one $k \geq 1$ such that $c_k \neq 0$. Under this orthogonal basis representation, the null hypothesis of zero treatment effect is equivalent to testing whether
\begin{equation} \label{eqn: simplehypothesis}
   H_{0,k}: c_k = 0 \quad \text{ vs } \quad H_{1,k}: c_k \neq 0,
\end{equation}
for all $k \geq 1$. This motivates us to use the projections $\{c_k\}_{k \geq 1}$ to test for the significance of the multivariate treatment effect. 

Consider the scalar projection of the centered process onto the eigenfunction $\bphi_k(t)$ as $\zeta_{i,k} := \langle \bX_i - \bmu,  \bphi_k\rangle$. Under~(\ref{eqn: meanofX}), $\mathbb{E}(\zeta_{i, k}) = g_i\,\langle \boldsymbol{\eta}, \boldsymbol{\phi}_k \rangle = g_i c_k \;$, and
    $\text{Var}(\zeta_{i,k}) = \langle \mathbf{\Xi} \bphi_k, \boldsymbol{\phi}_k \rangle$ for all $i=1,\dots,n$. Under the null hypothesis $H_0$, $\mathbb{E}(\zeta_{i, k}) = 0$ for all $i$; whereas under  $H_1$, at least for some $k \geq 1$, $\mathbb{E}(\zeta_{i, k}) = c_k \neq 0$ for all $i$ with $g_i=1$, i.e. for the treatment group.  Therefore, the random variable $\zeta_{i, k}$ can be used to test whether $c_k= 0$. A natural way to identify this is to measure the difference between the average of $\zeta_{i,k}$ between the treatment and the control group, i.e. $D_{k} := n_1^{-1}\sum_{i: g_i = 1} \zeta_{i, k} - n_0^{-1}\sum_{i: g_i = 0} \zeta_{i, k}$ for  $k \geq 1$
where $n_1 = \sum_{i=1}^n \mathbb{I}(g_i=1)$, and $n_0 = \sum_{i=1}^n \mathbb{I}(g_i=0)$ are the number of subjects in the treatment and the placebo group respectively. Note that $n_1 + n_0 = n$. Under the independence of the responses between the subjects, the projections are also independently distributed across $i$, hence $\text{Var}(D_k) = (1/n_1 + 1/n_0)\langle \mathbf{\Xi} \bphi_k, \boldsymbol{\phi}_k \rangle$. Thus, for every $k \geq 1$, $D_k$ can be used to test $H_{0,k}$ versus $H_{1,k}$.

To test the original null hypothesis $H_0$, we need to test $H_{0,k}$ versus $H_{1,k}$ for all $k \geq 1$. This means we have to test a countably infinite number of hypotheses to conclude globally about $H_0$, which is infeasible. For all practical purposes, we have to conduct simultaneous testing of $H_{0,k}$, $k=1, \dots,K$, for a large value of $K$.  Therefore, carrying out the projection-based inference procedure requires selection of orthogonal basis as well as the optimal number of orthogonal basis, $K$. Even after choosing $K$, we have to combine the inferences drawn from each of the simpler hypotheses to infer globally about $H_0$. 

Theoretically, any preset multivariate orthogonal basis functions such as multivariate version of Fourier basis, wavelets or Legendre basis will work. However the selection of truncation parameter $K$ under present basis becomes difficult, because i) there is no objective way to choose the optimal number of basis for a data, and typically that will require to test a large number of simpler hypotheses of the form $H_{0,k}$. ii) as the projections $\{\zeta_{i,k}\}_{k=1}^K$ are correlated among each other, it is not straightforward to handle to correlation among the difference vector $\{D_k\}_{k=1}^K$'s to further construct a test using $(D_1, \dots, D_K)^\top$, when $K$ is large. To avoid this, we choose a set of data-driven multivariate eigenbases from the covariance operator $\boldsymbol\Xi$, presented next. 

\subsection{Data-driven projection}

Assume that the symmetric non-negative covariance kernel of $\bX(t)$ is continuous, so that the linear covariance operator $\boldsymbol\Xi$ is compact, self-adjoint in $\mcL^2(\mcT)$ \citep[][chapter 7]{hutson2005applications}. By Hilbert-Schmidt theorem there exists a set of $q$-dimensional orthogonal basis functions (also known as eigenfunctions) $\{\boldsymbol\psi_{k}(\cdot) := (\psi^{(1)}_{k}(\cdot), \dots, \psi^{(q)}_{k}(\cdot))^\top \}_{k \geq 1} \in \mcL^2(\mcT)$ with 
$\langle \bpsi_k, \bpsi_{k'} \rangle =\mathbb{I}(k=k')$ for $k, k'\geq 1$ such that 
$$(\boldsymbol\Xi \bpsi_k)^{(\ell)}(t) =  \sum_{\ell^\prime=1}^q \langle \Sigma_{\ell\ell^\prime}(t, \cdot)\;,\psi_k^{(\ell^\prime)} \rangle = \lambda_k \bpsi_k^{(\ell)}(t) \qquad \ell=1,\dots, q,\;\;\;k=1,2,\dots, $$
where $\lambda_1 \geq \lambda_2 \geq \dots \geq 0$ are the ordered eigenvalues of $\boldsymbol\Xi$ with trace $\tr(\boldsymbol\Xi) = \sum_{k=1}^\infty \lambda_k < \infty$. By application of multivariate version of Mercer's theorem, the covariance kernel has the spectral decomposition
$
\boldsymbol \Sigma(t, t^\prime) = \sum_{k=1}^\infty \lambda_k \bpsi_k(t)\bpsi_k(t^\prime),
$ where the convergence is uniformly in $t \in \mcT$.
The eigenfunctions $\{\bpsi_k(t)\}_{k\geq 1}$ of the covariance operator $\boldsymbol \Xi$ will serve as our choice of orthogonal bases $\{\phi_k(t)\}$ (introduced in Section~\ref{sec: projectionbasedinference}) to carry out the data-driven projection-based test. Using these eigenbases, the latent process has the multivariate Kahrunen-Loeve (KL) representation,
\begin{equation}\label{eqn: multivriateKL}
    \bX_i(t) = \boldsymbol\mu(t) + \sum_{k=1}^\infty \xi_{i,k}\bpsi_k(t),
\end{equation}
where $\xi_{i,k} = \langle \bX_i - \boldsymbol\mu, \bpsi_k \rangle$ is the projection of $\bX_i(t)$ onto $q$-variate eigenfunction $\bpsi_k(t)$, known as multivariate functional principal component (mv-fPC) scores. We further use these scores $\{\xi_{i,k}\}_{i=1}^n$ to construct a test for $H_{0,k}$ for all $k \geq 1$. The advantages of using $\{\bpsi_k(t)\}_{k\geq 1}$ as our choice of orthogonal bases are multi-fold. First, the KL expansion provides a parsimonious representation of the latent process, and thus $\{\bpsi_k(t)\}_{k\geq 1}$ serves as a natural choice to represent the data. Second, one can objectively select the optimal number of eigenfunctions, $K$, by ensuring that the majority of the variation in the data is explained. In other words, if we truncate the above representation~(\ref{eqn: multivriateKL}) upto the first $K$ eigenfunctions, then $\bX_i^{(K)}(t) = \boldsymbol\mu(t) + \sum_{k=1}^K \xi_{i,k}\bpsi_k(t)$ is the best $K$-rank approximation of the infinite-dimensional stochastic process $\bX_i(t)$. Third, the mv-fPC scores $\{\xi_{i,k}\}_{k\geq 1}$ are different from $\{\zeta_{i,k}\}_{k\geq 1}$, the projections of $\bX_i(t) - \bmu(t)$ onto any arbitrary orthogonal bases $\{\boldsymbol\phi_k(t)\}_{k \geq 1}$, because the K-tuple $(\xi_{i,1}, \dots, \xi_{i,K})^\top$ serves as a key variation identifying substitute to the entire function $\bX_i(t)$, and they are uncorrelated among each other. Further, we get a simplified expression of the variance of $\xi_{i,k}$ as $\text{Var}(\xi_{i,k}) = \langle \mathbf{\Xi} \bpsi_k, \boldsymbol{\psi}_k \rangle = \lambda_k$.
Under~(\ref{eqn: meanofX}), the difference in the average of the mv-fPC scores, 
$$
 \mathcal{D}_{k} := n_1^{-1}\sum_{i: g_i = 1} \xi_{i, k} - n_0^{-1}\sum_{i: g_i = 0} \xi_{i, k} \qquad k=1,2,\dots,K
$$
has a mean $c^*_k = \langle \boldsymbol\eta, \boldsymbol\psi_k  \rangle$, the projection of treatment effect $\boldsymbol\eta(t)$ onto the eigenfunction $\boldsymbol\psi_k(t)$, and variance $\lambda_k(1/n_1 + 1/n_0)$. Moreover, $\mathcal{D}_k$ are uncorrelated among each other since the projections $\{\xi_{i,k}\}_{k=1}^K$ are uncorrelated across $k$.  Conditional on the finite truncation of eigenfunctions, $K$ (estimation of $K$ will be discussed in Section~\ref{sec: HTteststatistic}), the $K$-dimensional vector $\boldsymbol{\mathcal{D}} := (\mathcal{D}_1, \dots, \mathcal{D}_K)^{\top}$ has mean $(c^*_1, \dots, c^*_K)^\top$ and covariance $(1/n_1 + 1/n_0)\text{diag}(\lambda_1, \dots, \lambda_K)$. Note that under $H_0$, $(c^*_1, \dots, c^*_K) \equiv 0$. Thus, we lay out a framework for detecting difference of the mean between two groups of multivariate functional data in terms of the difference of $K$-dimensional data-driven scores between the two groups obtained by multivariate functional principal component analysis (fPCA). We formally present our score based test in the next section.

\section{Projection-based Multivariate FPCA score based test} \label{sec: HTteststatistic}

The true eigenfunctions and the associated mv-fPC scores are unknown, and they need to be estimated from the data. We briefly describe the estimation strategy of eigencomponents here. Let $\widehat{\boldsymbol\mu}(t)$ and $\widehat{\bSigma}(t,t^\prime)$ are consistent estimator of the multivariate mean and the covariance function. See \cite{chiou2014multivariate, happ2018multivariate, li2020fast} for different smoothing methods applied for estimation of covariance function for multivariate functional data. A consistent estimator of $\widehat{\boldsymbol\phi}_k(t)$ is obtained by spectral decomposition of the estimated covariance, i.e. $\widehat{\bSigma}(t, t^\prime) = \sum_{k} \widehat{\lambda}_k \widehat{\bpsi}_k(t)\widehat{\bpsi}_k(t^\prime)$. The number of optimal eigenfunctions, $K$, can be obtained either by setting a pre-specified proportion of percentage of variation explained (PVE), or Akaike information criterion (AIC). For densely observed functional data one can obtain a consistent estimator of the mv-fPC scores by computing $\widehat{\xi}_{ik} = \langle \bX_i - \widehat{\boldsymbol\mu}, \widehat{\boldsymbol\phi}_k \rangle$,   However, for sparse data, we do not observe the entire function $\bX_i(t)$. In this case we propose to consider $\widehat{\xi}_{ik} = \langle \widehat{\bX}_i - \widehat{\boldsymbol\mu}, \widehat{\boldsymbol\phi}_k \rangle, k=1, \dots, K$, as an estimator of the scores, where $\widehat{\bX}_i(t)$ is the best linear unbiased predictor of the unobserved trajectory $\bX_i(t)$ under a working Gaussian assumption \citep{li2020fast}. A detailed description of the estimation method of covariance function as well as an explicit formula for the $\widehat{\xi}_{ik}$ is provided in Section \ref{sec: mface_estimation_of_eigenfun} of the supplementary material.
 
 Let $\widehat{\boldsymbol\xi}_i = (\widehat{\xi}_{i,1}, \dots, \widehat{\xi}_{i,K})^\top$ be the estimated $K$-dimensional multivariate fPC scores for the $i$th subject. Define, $\widehat{\boldsymbol\xi}_{1+} = n_1^{-1}\sum_{i: g_i = 1} \widehat{\boldsymbol\xi}_{i}$, and the $\widehat{\boldsymbol\xi}_{0+} = n_0^{-1}\sum_{i: g_i = 0} \widehat{\boldsymbol\xi}_{i}$ are the average of the estimated scores for the treatment and placebo group. Similarly define, $\widehat{\boldsymbol\Lambda}_1 = (n_1-1)^{-1}\sum_{i: g_i = 1} (\widehat{\boldsymbol\xi}_{i} - \widehat{\boldsymbol\xi}_{1+})(\widehat{\boldsymbol\xi}_{i} - \widehat{\boldsymbol\xi}_{1+})^\top$, and $\widehat{\boldsymbol\Lambda}_0 = (n_0-1)^{-1}\sum_{i: g_i = 0} (\widehat{\boldsymbol\xi}_{i} - \widehat{\boldsymbol\xi}_{0+})(\widehat{\boldsymbol\xi}_{i} - \widehat{\boldsymbol\xi}_{0+})^\top$ are the sample variance of the BLUP estimated scores for the two group. Further, define the pooled sample covariance as $\widehat{\boldsymbol\Lambda} = \{(n_1-1)\widehat{\boldsymbol\Lambda}_1 + (n_0-1)\widehat{\boldsymbol\Lambda}_0\}/(n-2)$. To test for $H_0$, a Hotelling $T$-squared statistic \citep[Equation 10, Chapter 6][]{muirhead2009aspects} using the mv-fPC scores can be constructed as
   \begin{equation} \label{eqn: HotellingT2}
     T_{n}  = \frac{n_1n_0 }{n_1+n_0} \widehat{\boldsymbol{\mathcal{D}}}^\top \widehat{\boldsymbol\Lambda}^{-1} \widehat{\boldsymbol{\mathcal{D}}},
\end{equation}
where $\widehat{\boldsymbol{\mathcal{D}}} = \widehat{\boldsymbol\xi}_{1+} - \widehat{\boldsymbol\xi}_{0+}$. 
Under the null hypothesis, $T_n$ is approximately chi-square distributed with $K$ degrees of freedom, for large $n$. However, for fixed sample size $n$, to account of the variation due to estimating the true covariance of the fPC scores by its sample version, we relate the Hotelling $T$-squared distribution to the $F$-distribution \citep[Theorem 5.9, ][]{hardle2019applied} under a working Gaussian assumption on the mv-fPC scores.  
Our test rule rejects $H_0$ at a specified significance level $\alpha \in (0,1)$ if 
\begin{equation} \label{eqn: HotellingTestRule_eqvar}
    T_n > \frac{(n-2)K}{(n - K - 1)}F_\alpha( K ,n - K - 1),
\end{equation}
where $K$ is the dimension of mv-fPC scores estimated from the data, and  $F_\alpha(a,b)$ is the $100(1-\alpha)\%$ quantile of $F$-distribution with $a$ and $b$ degrees of freedom. The next proposition theoretically confirms that the test maintains the type-I error.

\begin{prop} \label{thm: type1error}
Assume that model~(\ref{eqn: model}) for the observed response $\{\bY_{ij}: j=1, \dots, m_i\}_{i=1}^n$ is true, and $\sup_i m_i < \infty$. Further assume that the mean functions and eigencomponents are estimated consistently, i.e. $\lVert \widehat{\bmu} - \bmu\rVert = o_p(1)$, $\lVert \widehat{\boldsymbol\psi}_k - \boldsymbol\psi_k \rVert = o_p(1)$, $\lVert \widehat{\lambda}_k - \lambda_k\rVert = o_p(1)$ for all $k=1, \dots, K$, and  $\lVert \widehat{\tau}^2_\ell - \tau_\ell^2\rVert = o_p(1)$ for all $\ell =1, \dots, q$, and that $\lim_{n \to \infty} n_1/n \to w \in (0,1)$. Then, conditional on the truncation parameter $K$, under the null hypothesis for any $\alpha \in (0,1)$
$$
\mathbb{P}\left(T_n > \frac{(n-2)K}{(n - K - 1)}F_\alpha( K ,n - K - 1) \Bigm| H_0 \text{ is true}\right) \leq \alpha \qquad \text{as } n\to \infty.
$$
\end{prop}
The assumptions made for the above proposition is quite standard in fDA literature. The consistency of the eigencomponents from mv-fPCA are established in \cite{chiou2014multivariate}. The proof of the proposition follows similar to the proof of Theorem 1 of \cite{wang2021two} for the univariate case and the fact that dimension of multivariate response $q$ is finite, and is omitted here to avoid redundancy. 

The proposition is presented conditional on the number of estimated eigenfunctions $K$. It is important to note that the number of estimated eigenfunctions $K$ increases with $n$. However, the rate of growth at which $K$ grows with $n$ is much slower. Keeping the $K$ fixed does not affect the size of the test, nonetheless it may affect the power. Specifically, consider a hypothetical situation where the true eigenbasis of the covariance is truncated to $\{\bpsi_k(t)\}_{k=1}^K$, and the projection of $\boldsymbol\eta(t)$ along these $K$ leading directions is null, but it is significantly different from zero along the $(K+1)$th direction, $\bpsi_{K+1}(t)$. Testing the null hypothesis along the leading $K$ eigenfunctions will have no power. However, setting a large PVE (say $99\%$ or $99.9\%$) will lead to selection of higher number of eigenfunctions, thereby reduce the chance of falling in this untoward situation.

\subsubsection*{Permutation-based approximation of null distribution:} For small sample sizes (say $n < 50$), the null distribution of the Hotelling $T$-squared statistic can also be approximated by permutation test, instead of using the $F$-distribution. The steps of the permutation test are as follows. 
\begin{enumerate}
    \item Split the entire sample of mv-fPC scores $\{\widehat{\boldsymbol\xi}_{i}\}_{i=1}^n$ for all subjects randomly into two groups of size $n_1$ and $n_0$. Assign $g_i=1$ for the first group, and $g_i=0$ for the second group.
    \item Compute the value of the test-statistic using the new permuted sample using equation~(\ref{eqn: HotellingT2}).
    \item Repeat step 1-2 $B$ times to get samples $\{{T}_{n,b}\}_{b=1}^B$ from the null distribution of the test-statistic, for a large value of $B$, say $5000$. 
    \item The p-value is computed as $B^{-1}\sum_{b=1}^B \mathbb{I}({T}_{n,b} > {T}_{n})$ to conclude about $H_0$.
\end{enumerate}

\subsubsection*{Test-statistic under unequal variance:} The Hotelling $T$-squared statistic constructed in~(\ref{eqn: HotellingT2}) is under the assumptions that the variance of mv-fPC scores of the two groups are same, which is primarily driven by the assumption we made that the covariance of $\bX_i(t)$ is same for all $i=1,\dots,n$, irrespective of the group in which the subject $i$ belongs to. However, this assumption may not always hold. In that case, we can carry out the test by approximating the statistic using Satterthwaite approximation assuming unequal variance, \citep{nel1986solution} as, 
\begin{equation*}
    T_{n,UV} = \widehat{\boldsymbol{\mathcal{D}}}^\top(n_1^{-1}\widehat{\boldsymbol\Lambda_1} + n_0^{-1}\widehat{\boldsymbol\Lambda_0})^{-1}\widehat{\boldsymbol{\mathcal{D}}}.
\end{equation*}
We reject $H_0$ at a level $\alpha \in (0,1)$ if 
\begin{equation}\label{eqn: HotellingTestRule_uneqvar}
    T_{n, UV} > \frac{fK}{(f - K - 1)}F_\alpha( K ,f - K - 1),
\end{equation}
where $f$ is the effective degrees of freedom. The exact formula for $f$ is given in \citet[][page 12]{nel1986solution}. 

\subsubsection*{Inference for outcome-specific mean function:} After globally concluding about the null hypothesis, if it were rejected, the natural question of interest would be to identify which of the $q$ outcomes are significantly different between the treatment and the placebo group. One can answer this question by individually inspecting the estimated treatment effect $\widehat{\eta}^{(\ell)}(t)$, $\ell=1, \dots, q$, and obtain a bootstrap standard errors for $\widehat{\eta}^{(\ell)}(t)$. See \cite{park2018simple} for the bootstrapping mechanism of subjects for functional data. After obtaining the bootstrap standard errors, $100(1-\alpha)\%$ simultaneous confidence band for the true treatment effects ${\eta}^{(\ell)}(t)$ can be obtained by algorithm 2 of \cite{cui2022fast}. To adjust for type-I error in the multiple testing of $q$-components of treatment effects separately, we set $\alpha^* = \alpha/q$ to construct the individual confidence band. 
\subsubsection*{Extension to more than two groups:}
When the number of groups, $G$, is more than two, testing equality of the multivariate mean functions between $G$ groups translates to testing whether the  $K$-dimensional mv-fPC scores of the subjects in the $G$ groups, $\{\widehat{\boldsymbol\xi}_i : g_i = r\}_{r=1}^G$ have the same mean. This resembles to a multivariate analysis of variance (MANOVA) problem involving the mv-fPC scores. Define, $n_r = \sum_{i=1}^n I(g_i = r)$ as the group frequency, $\widehat{\boldsymbol\xi}_{r+} = n_r^{-1}\sum_{i: g_i = r} \widehat{\boldsymbol\xi}_{i}$ as the group mean, and $\widehat{\boldsymbol\Lambda}_r = (n_r-1)^{-1}\sum_{i: g_i = r} (\widehat{\boldsymbol\xi}_{i} - \widehat{\boldsymbol\xi}_{r+})(\widehat{\boldsymbol\xi}_{i} - \widehat{\boldsymbol\xi}_{r+})^\top$ to be the covariance for the $r$th group respectively, $r=1, \dots, G$. Also, denote by $\widehat{\boldsymbol\xi}_{++} = n^{-1}\sum_{i=1}^n \widehat{\boldsymbol\xi}_{i}$ as the overall mean of the mv-fPC scores. To carry out the MANOVA, we construct the {\em among groups error} matrix as $\mathbf{Q}_H = \sum_{r=1}^G n_r (\widehat{\boldsymbol\xi}_{r+}-\widehat{\boldsymbol\xi}_{++})(\widehat{\boldsymbol\xi}_{r+}-\widehat{\boldsymbol\xi}_{++})^\top$, and the {\em among units error} matrix as $\mathbf{Q}_E = \sum_{r=1}^G (n_r-1)\widehat{\boldsymbol\Lambda}_r$. There are several test-statistics that are constructed based on the $\mathbf{Q}_H$ and $\mathbf{Q}_E$ such as Wilk's Lambda, Lawley-Hotelling trace, Pillai's trace, to name some \citep{johnson2002applied}. We present the form of Lawley-Hotelling trace statistic which reject $H_0$ large values of 
\begin{equation*}
    T_{LH} = \tr(\mathbf{Q}_H\mathbf{Q}_E^{-1}).
\end{equation*}
Under the null hypothesis, $T_{LH}$ is approximately $\chi^2$ distributed with $K(G-1)$ degrees of freedom. Note that $K$ is the dimension of the mv-fPC scores estimated from the data. None of the above statistic named above is shown to be superior to the others in terms of power. All of them reduce to the Hotelling $T$-squared statistic in~(\ref{eqn: HotellingT2}) for $G=2$.

\section{Numerical studies} \label{sec: mface_testing_simstudy}

In this section, we numerically investigate the finite-sample type-1 error and the power of our proposed test to detect the departure from null hypothesis. To accomplish that we generate three-dimensional (i.e. $q=3$) functional data by the following mechanism,
\begin{equation*}
    Y_{ij}^{(\ell)} = \mu^{(\ell)}(t_{ij}) + g_i\eta^{(\ell)}(t_{ij})  + \sum_{k=1}^3 \xi_{ik} \psi_k^{(\ell)}(t_{ij}) + \epsilon^{(\ell)}_{ij},
\end{equation*}
for $\ell=1,2,3, \;j=1, \dots, m_{i}$, and $i=1, \dots, n$. The observation points $t_{ij}$ are uniformly sampled from $[0,1]$ for all $i,j$.  The mean vector is taken as $\mu^{(1)}(t) = 5\sin(2\pi t)$, $\mu^{(2)}(t) = 5\cos(2\pi t)$, and $\mu^{(3)}(t) = 5(t-1)^2$, and the treatment effect is considered as $\eta^{(\ell)}(t) =  5\delta(t/4-0.5)^3$ for $\ell=1,2,3$. The parameter $\delta$ controls the departure from the null hypothesis, with $\delta = 0$, implying that $H_0$ is true. The group indicator $g_i$ are randomly selected from $\{0,1\}$ for each $i=1,\dots,n$. For the random components, the fPC scores $\xi_{i,k}$ has a mean zero, and variance $\lambda_k$ for $k=1, 2, 3$ with $\lambda_1 = 6, \lambda_2 = 3$, and $\lambda_3 = 1.5$. The orthonormal eigenfunctions are taken as $\bpsi_1(t) = \sqrt{2/3}\left[\sin(2\pi t), \cos(4\pi t), \sin(4\pi t)\right]^\top$, $\bpsi_2(t) = \sqrt{2/3}\left[\sin(\pi t/2), \sin(3\pi t/2), \sin(5\pi t/2)\right]^\top$, and $\bpsi_3(t) = \sqrt{2/3}\left[\sin(\pi t), \sin(2\pi t), \sin(3\pi t)\right]^\top$. The measurement errors $\epsilon_{ij}^{(\ell)} \sim (0,\sigma_e^2)$ with $\sigma_e = 0.2$, for all $i,j$, and $\ell$.

We consider a factorial design with three important factors, the sample size $n$, the sparsity level of the observed functions, and the distribution of the fPC scores. Five different values of the sample size $n$ is considered, $n=50, 70, 100, 200$, and $300$, to demonstrate the performance of the test for small to large sample sizes. We choose three different levels of sparsity; which are based on the number of times the functions are observed, i.e. {\em high}: $m_{i} \in \{4,5,6,7\}$, {\em medium}: $m_{i} \in \{8,9,10,11,12\}$, and {\em low}: $m_{i} \in \{15, 16, 17, 18, 19, 20\}$ for all $i$. For a chosen $m_{i}$, the observation points $t_{ij}$ are randomly chosen from a equidistant grid of $51$ points in $[0,1]$ without replacement. For the last factor, the scores $\xi_{i,k}$ are generated from either {\em Gaussian} distribution or a {\em non-Gaussian} distribution, in which case, they are generated from a mixture of two normals, $N(\sqrt{\lambda_k/2}, \lambda_k/2)$ with probability $0.5$, and $N(-\sqrt{\lambda_k/2}, \lambda_k/2)$ with probability $0.5$, $k=1,2,3$.

\subsection{Computational details}
For each component $\ell=1, 2, 3$, we estimate the common mean function $\mu^{(\ell)}(t)$ and the treatment effect $\eta^{(\ell)}(t)$ using the \verb|gam()| function in \verb|mgcv| package \citep{wood2011fast} under a working independence assumption to obtain the residual as $E_{ij}^{(\ell)} = Y_{ij}^{(\ell)} - \widehat{\mu}^{(\ell)}(t_{ij}^{(\ell)}) - g_i\widehat{\eta}^{(\ell)}(t_{ij}^{(\ell)})$ for all $i,j$, and $\ell$. The number of knots are taken as $10$, and the they are placed uniformly. In the second step, we obtain the estimated eigenvalues $\widehat{\lambda_k}$, and the eigenfunctions $\widehat{\bpsi_k}(t)$ by applying the multivariate functional principal component analysis (mv-fPCA) on the residuals $\{E_{ij}^{(\ell)}\}_{i,j,\ell}$, using \verb|mface.sparse()| function in the R package \verb|mfaces| \citep{mfaces}. The number of eigenfunctions $K$ are chosen by pre-specified PVE of $99\%$. At the third step, the mv-fPC scores $\widehat{\boldsymbol\xi}_i = (\widehat{\xi}_{i,1}, \dots, \widehat{\xi}_{i,K})^\top$ are estimated by the method described in section~\ref{sec: HTteststatistic} for all $i=1,\dots,n$. Finally, the estimated scores $\{\widehat{\boldsymbol\xi}_i : g_i =1\}_{i=1}^n$ and $\{\widehat{\boldsymbol\xi}_i : g_i =0\}_{i=1}^n$ for the two groups are tested for equality using the using \verb|hotelling.test()| function in R package \verb|Hotelling| \citep{hotelling} to conclude about the null hypothesis $H_0$, based on a single p-value.

The computation cost of our testing procedure is primarily driven by the complexity in carrying out the multivariate fPCA, which requires $\mathcal{O}(n \max \{q^3 m, q^2m^2\})$ operations. Here $m = \max_i m_i$. Therefore, when the number of observations for each subject, $m_i$, is finite, the complexity of the test grows at a cubic rate with the dimension of the multivariate response. Table~\ref{tab: computation time} documents the median computational time (in seconds) for one replication across different sample sizes and sparsity levels. 
\begin{table}
	\centering
	\caption{Computation time (in seconds)  for one replication}
	\scalebox{0.8}{
	\begin{tabular}{cccc}
		\hline	\hline
		& \textit{high} & \textit{medium} & \textit{low} \\ 
		\hline
	$ n = 100$ &	 14.26 & 15.30 & 18.64  \\ 
		$ n = 200$&	27.08 & 28.33 & 38.72  \\ 
		\hline	\hline
	\end{tabular}}\label{tab: computation time}
\end{table}

\subsection{Assessing performance of the test}
\begin{table}
	\centering
	\caption{The empirical type I error rates of the proposed test based on 10,000 simulations. Standard errors are presented in parentheses.\\}  
	\scalebox{0.95}{
		\begin{tabular}{cc cccc}
		\toprule
		\multicolumn{6}{c}{score distribution: {\em Gaussian}} \\ 
			\hline	\hline
			\multicolumn{6}{c}{sparsity: {\em high}  $\quad( m_i \sim \{4, \ldots, 7\})$} \\	\hline
			& & $\alpha = 0.01$ & 	$\alpha = 0.05$ & 	$\alpha = 0.10$ & 	$\alpha = 0.15$ \\ 
			\hline	\hline
			$n = $	50	 &&  0.010  (0.001) & 0.055  (0.002) & 0.108  (0.003) & 0.163  (0.004) \\	
	$n = $	70	 &&  0.012  (0.001) & 0.051  (0.002) & 0.106  (0.003) & 0.156 (0.004) \\  
	$n = $	100	 &&  0.010  (0.001) & 0.051  (0.002) & 0.104 (0.003) & 0.160  (0.004) \\ 
	$n = $	200	 && 0.009  (0.001) & 0.049  (0.002) & 0.101  (0.003) & 0.153  (0.004) \\
	$n = $	300	 && 0.011  (0.001) & 0.049  (0.002) & 0.100  (0.003) & 0.149  (0.004) \\
			\hline	\hline
					\multicolumn{6}{c}{sparsity: {\em medium}  $\quad( m_i \sim \{8, \ldots, 12\})$} \\	\hline
			& & $\alpha = 0.01$ & 	$\alpha = 0.05$ & 	$\alpha = 0.10$ & 	$\alpha = 0.15$ \\ 
			\hline	\hline
			$n = $	50	 &&  0.009  (0.001) & 0.051  (0.002) & 0.103  (0.003) & 0.159  (0.004) \\	
	$n = $	70	 &&  0.010  (0.001) & 0.053  (0.002) & 0.101  (0.003) & 0.152 (0.004) \\  
	$n = $	100	 &&  0.009  (0.001) & 0.047  (0.002) & 0.097 (0.003) & 0.152  (0.004) \\ 
	$n = $	200	 && 0.009  (0.001) & 0.048  (0.002) & 0.097  (0.003) & 0.143  (0.004) \\
	$n = $	300	 && 0.009  (0.001) & 0.047  (0.002) & 0.094  (0.003) & 0.148  (0.004) \\
	\hline \hline
				\multicolumn{6}{c}{sparsity: {\em low}  $\quad( m_i \sim \{15, \ldots, 20\})$} \\	\hline
			& & $\alpha = 0.01$ & 	$\alpha = 0.05$ & 	$\alpha = 0.10$ & 	$\alpha = 0.15$ \\ 
			\hline	\hline
			$n = $	50	 &&  0.009  (0.001) & 0.051  (0.002) & 0.100  (0.003) & 0.151  (0.004) \\	
	$n = $	70	 &&  0.010  (0.001) & 0.050  (0.002) & 0.098  (0.003) & 0.144 (0.004) \\  
	$n = $	100	 &&  0.010  (0.001) & 0.047  (0.002) & 0.096 (0.003) & 0.148  (0.004) \\ 
	$n = $	200	 && 0.010  (0.001) & 0.048  (0.002) & 0.099  (0.003) & 0.146  (0.004) \\
	$n = $	300	 && 0.010  (0.001) & 0.050  (0.002) & 0.099  (0.003) & 0.152  (0.004) \\
	\bottomrule
	\toprule
			\hline \hline 
		\multicolumn{6}{c}{score distribution: {\em Mixture of Gaussian}} \\ 
			\hline	\hline
			\multicolumn{6}{c}{sparsity: {\em high}  $\quad( m_i \sim \{4, \ldots, 7\})$} \\	\hline
			& & $\alpha = 0.01$ & 	$\alpha = 0.05$ & 	$\alpha = 0.10$ & 	$\alpha = 0.15$ \\ 
			\hline	\hline
			$n = $	50	 &&  0.010  (0.001) & 0.053  (0.002) & 0.112  (0.003) & 0.164  (0.004) \\	
	$n = $	70	 &&  0.012  (0.001) & 0.059  (0.002) & 0.112  (0.003) & 0.163 (0.004) \\  
	$n = $	100	 &&  0.011  (0.001) & 0.052  (0.002) & 0.105 (0.003) & 0.158  (0.004) \\ 
	$n = $	200	 && 0.008  (0.001) & 0.046  (0.002) & 0.095  (0.003) & 0.148 (0.004) \\
	$n = $	300	 && 0.009  (0.001) & 0.046  (0.002) & 0.097  (0.003) & 0.146 (0.004) \\
			\hline	\hline
					\multicolumn{6}{c}{sparsity: {\em medium}  $\quad( m_i \sim \{8, \ldots, 12\})$} \\	\hline
			& & $\alpha = 0.01$ & 	$\alpha = 0.05$ & 	$\alpha = 0.10$ & 	$\alpha = 0.15$ \\ 
			\hline	\hline
			$n = $	50	 &&  0.011  (0.001) & 0.056  (0.002) & 0.110  (0.003) & 0.161  (0.004) \\	
	$n = $	70	 &&  0.010  (0.001) & 0.052  (0.002) & 0.104  (0.003) & 0.154 (0.004) \\  
	$n = $	100	 &&  0.011 (0.001) & 0.052  (0.002) & 0.101 (0.003) & 0.154 (0.004) \\ 
	$n = $	200	 && 0.010  (0.001) & 0.050  (0.002) & 0.104  (0.003) & 0.155  (0.004) \\
	$n = $	300	 && 0.009  (0.001) & 0.048  (0.002) & 0.100  (0.003) & 0.153  (0.004) \\
	\hline \hline
				\multicolumn{6}{c}{sparsity: {\em low}  $\quad( m_i \sim \{15, \ldots, 20\})$} \\	\hline
			& & $\alpha = 0.01$ & 	$\alpha = 0.05$ & 	$\alpha = 0.10$ & 	$\alpha = 0.15$ \\ 
			\hline	\hline
			$n = $	50	 &&  0.011  (0.001) & 0.052  (0.002) & 0.104  (0.003) & 0.152  (0.004) \\	
	$n = $	70	 &&  0.011  (0.001) & 0.050  (0.002) & 0.100  (0.003) & 0.154 (0.004) \\  
	$n = $	100	 &&  0.010 (0.001) & 0.050  (0.002) & 0.104 (0.003) & 0.156 (0.004) \\ 
	$n = $	200	 && 0.011  (0.001) & 0.049  (0.002) & 0.099  (0.003) & 0.150  (0.004) \\
	$n = $	300	 && 0.009  (0.001) & 0.046  (0.002) & 0.094  (0.003) & 0.146 (0.004) \\
	\hline \hline
		\end{tabular}
	} \label{tab: size_multi}
\end{table}

\subsubsection{Size properties}
The empirical type-I error of the proposed test is presented in Table~\ref{tab: size_multi} for nominal levels $\alpha = 0.01,0.05,0.1,0.15$  across different sample sizes, sparsity levels for both when the scores are generated from the {\em Gaussian} and {\em non-Gaussian} distribution. The numbers are obtained by using the test rule~(\ref{eqn: HotellingTestRule_eqvar}). The estimates and the standard errors (in the parenthesis) are obtained based on $10,000$ simulations. Except for few cases of small sample sizes ($n = 50, 70$), and {\em high} sparsity levels, the empirical size is within twice standard error of the nominal level $\alpha$, which implies that our test maintains the type-I error quite well. Even for $n = 50$ case, we see that the inflation of type-I error comes under control as the sparsity decreases from {\em high} to {\em low}. This suggests that the proposed test can be conducted with a small samples in each group provided the the sparsity level of observed response is not too high. The slight inflation of the size can be attributed to the poor estimation of the eigenfunctions due to small sample sizes and/or higher sparsity level of the observed functions. It is interesting to see that the size of the test is also under control even when the scores are generated from a {\em non-gaussian} distribution, reflecting the robustness of our test to the deviation from normality.

\subsubsection{Power}

\begin{figure}
    \centering
    \subfloat[Gaussian data]{\includegraphics[scale=0.26]{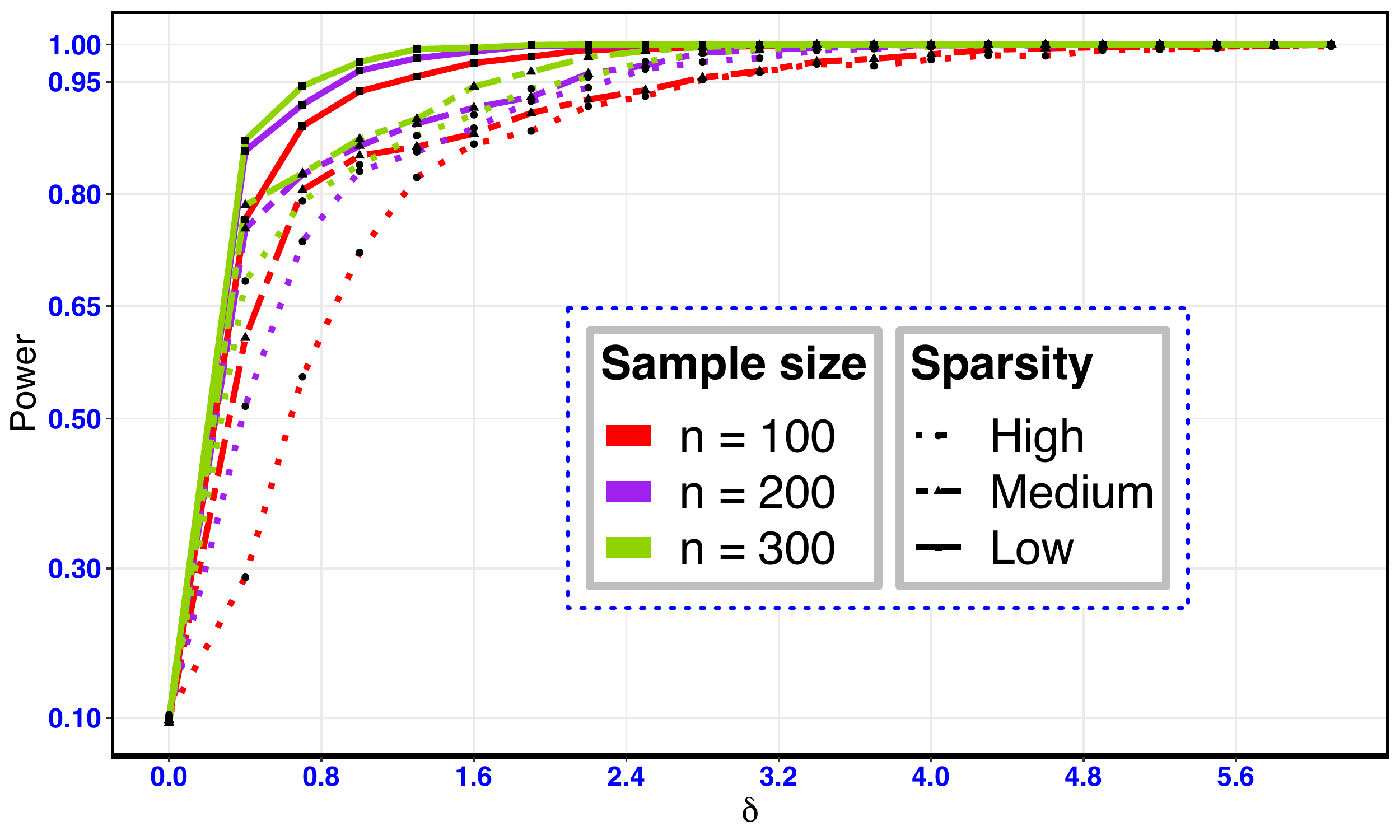}}
        \subfloat[Non-Gaussian data]{\includegraphics[scale=0.26]{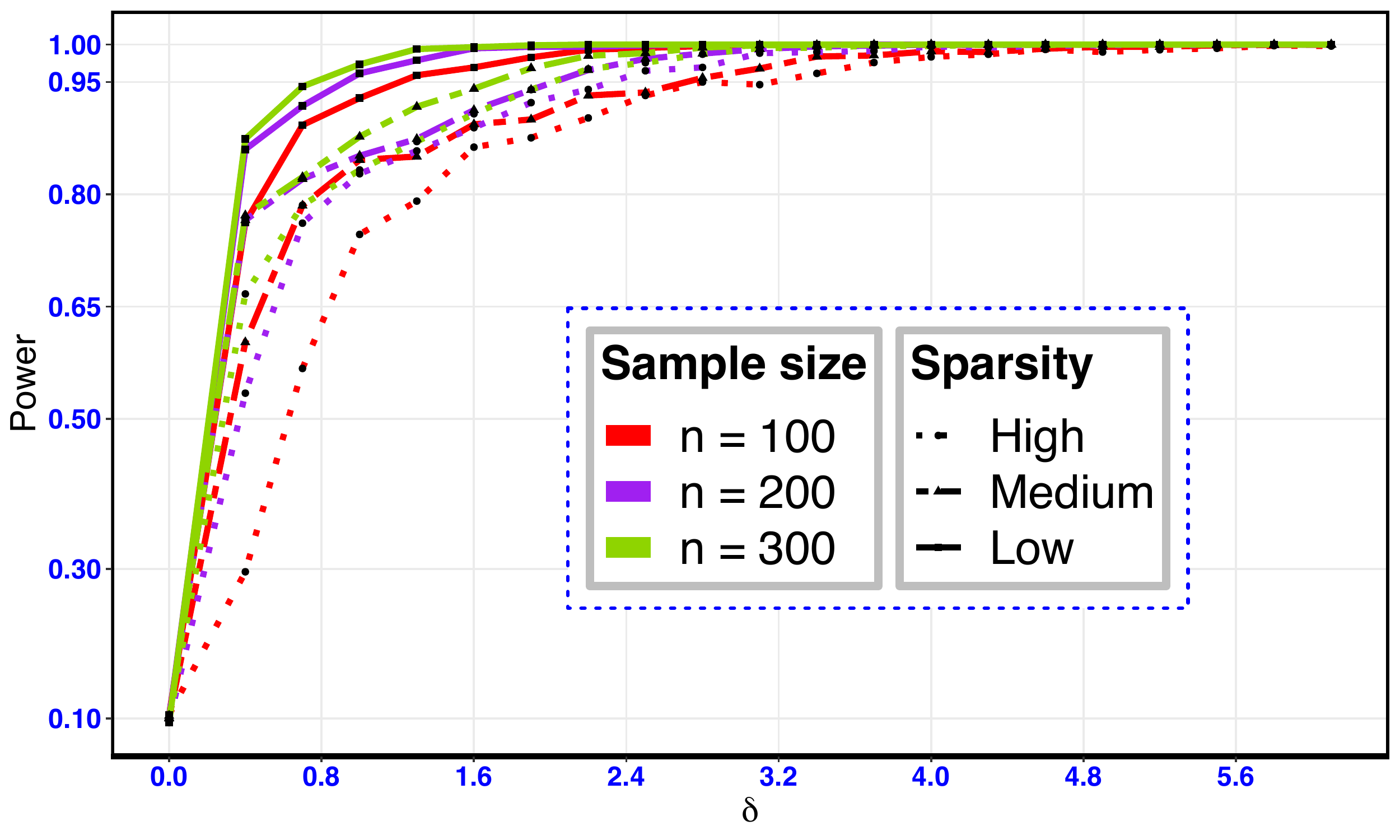}}
    \cprotect\caption{The empirical power of the proposed test as a function of $\delta$ (in the x-axis). (a) The left panel corresponds to case of Gaussian score, and (b) the right panel is for the non-Gaussian case. The quantity $\delta$ measures the departure from $H_0$.}
    \label{fig: power_multi}
\end{figure}

Fix the significance level of the test at $\alpha = 0.1$. Figure \ref{fig: power_multi}(a) shows the power curve of the test proposed test as function of $\delta$ for different sample sizes and the sparsity level of the observed functions when scores are {\em Gaussian}, and Figure \ref{fig: power_multi}(b) shows the same for the {\em non-Gaussian} scores. As expected, the power of the test increases with the sample size. Moreover, as the sparsity level decreases, the the power increases quickly for fixed sample size. 
The power curves in left panel of Figure \ref{fig: power_multi} in comparison to the right  panel demonstrate the robustness of the proposed test since it exhibits similar power when the data distribution deviates from normality.

\begin{figure}
    \centering
    \subfloat[Gaussian data]{\includegraphics[scale=0.3]{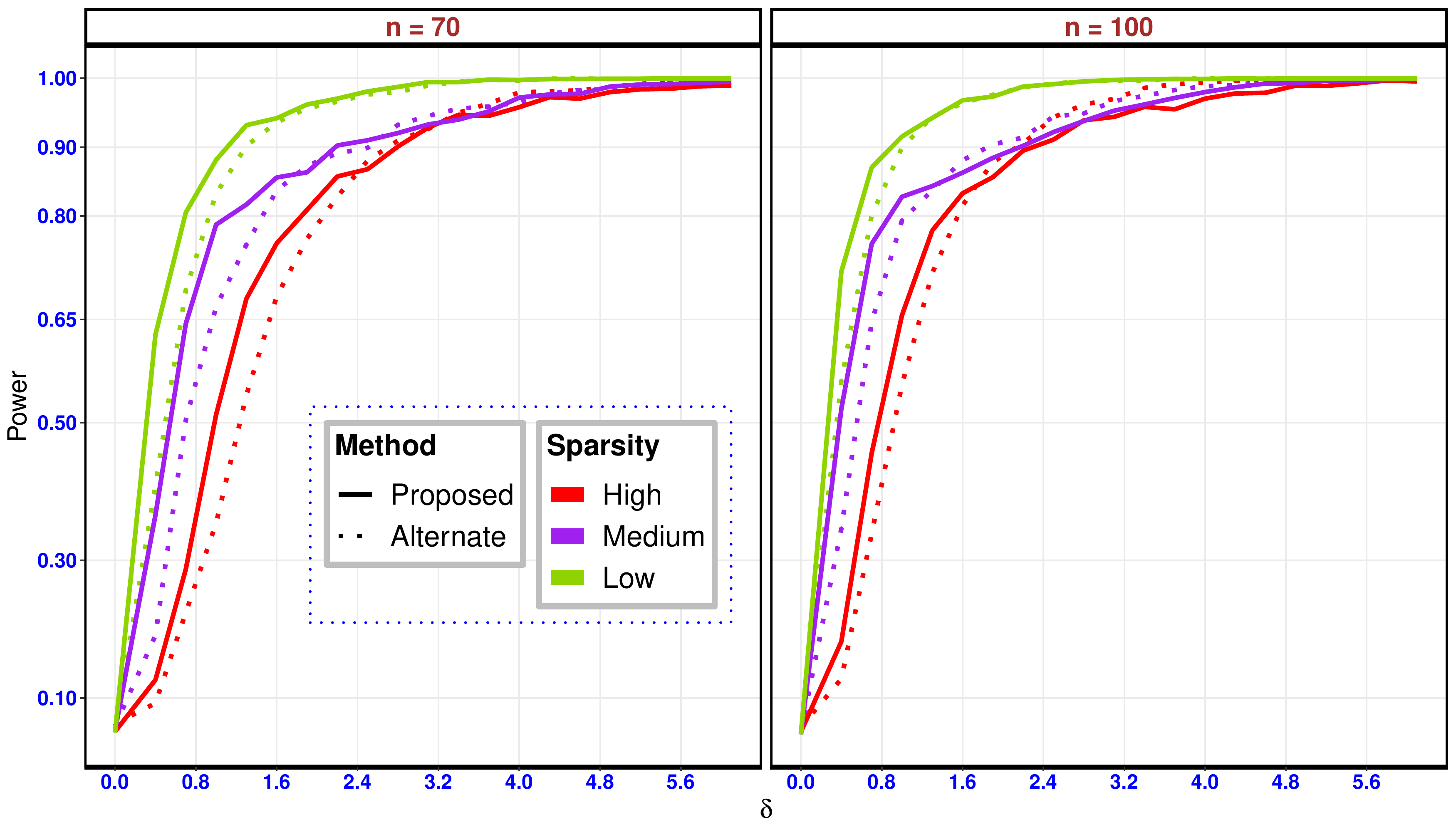}}\\
        \subfloat[Non-Gaussian data]{\includegraphics[scale=0.3]{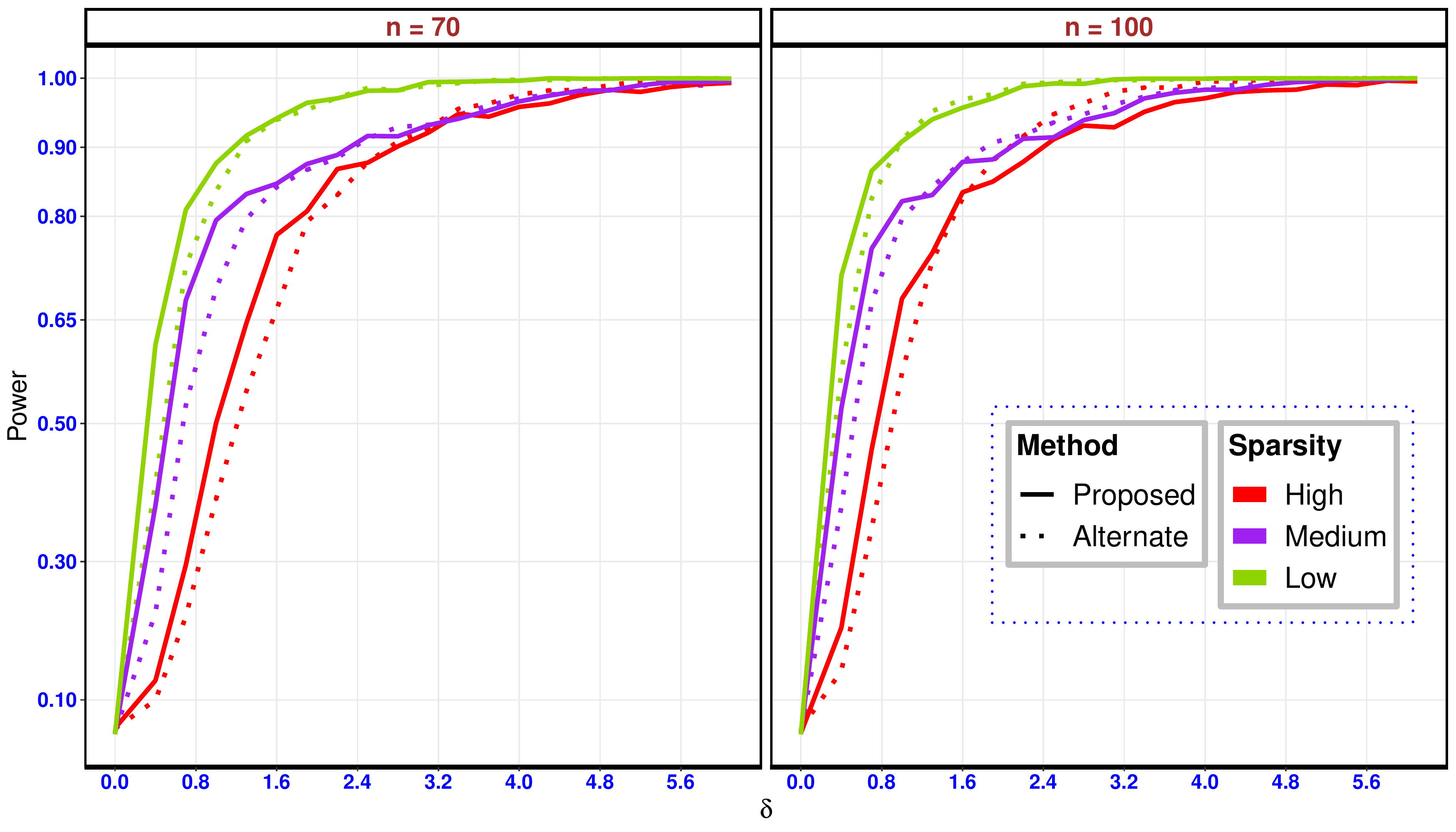}}
    \cprotect\caption{The empirical power of the proposed test as a function of $\delta$ (in the x-axis) compared to the multiple testing procedure proposed by \cite{pomann2016two}. (a) The upper panel corresponds to case of Gaussian score, and (b) the lower panel is for the non-Gaussian case. The quantity $\delta$ measures the departure from $H_0$.}
    \label{fig: power_compare}
\end{figure}

We further compare the performance of our test with the inference procedure proposed by \cite{pomann2016two} that conducts multiple two-sample univariate
tests using the fPC scores for each $k=1,\dots,K$, combined with a multiple comparison adjustment using Bonferrroni's correction. Although this procedure is primarily developed for univariate fD, we replicate it for the multivariate fD case. As presented in Figure~\ref{fig: power_compare} we can see that our proposed test exhibits superior power than it primary competitor, across all the sparsity levels as well as for both the Gaussian (Figure~\ref{fig: power_compare}(a)), and the non-Gaussian case (Figure~\ref{fig: power_compare}(b)), especially when the deviation from the null hypothesis is small, and the sample size is not large. This demonstrates that when the detection of the departure from null is potentially difficult, our method is more powerful than the alternative. This is primarily due to the conservative nature of the multiple testing procedure due to adjusting the type I error using the Bonferroni's correction. 



\section{Analysis of TOMMORROW study} \label{sec: mface_testing_azillectstudy}

\subsection{Details on the study}
The data was collected from a phase 3 multi-centre, randomized placebo-controlled double blind parallel-group study of cognitive healthy individuals from the affiliate and private research clinics in the USA, aged between 65 and 83 years, who were at the high-risk of developing MCI due to Alzheimer's disease. The participants were randomly assigned 1:1 to receive the study drug pioglitazone or placebo tablets with identical appearance orally. The participants were assessed at baseline, and every 6 month intervals at a clinic visit, where participants' clinical and cognitive status were evaluated through questionnaires filled up by participant and the study partner, who is aware of the participant's cognitive status. The overall duration of the study is about 5 years. The final data consists of 1947 high-risk individuals with $985$ subjects in the pioglitazone, and $962$ subjects in the placebo group. See \cite{burns2021safety} for a more detailed description on the design of the study.

\begin{table}
\begin{tabular}{ll}
\hline
\multicolumn{1}{c}{\textbf{Cognitive   Domain}} & \multicolumn{1}{c}{\textbf{Tests}} \\ \hline
Episodic   memory & \begin{tabular}[c]{@{}l@{}}CVLT Short delay, \\ CVLT Long delay, \\ BVMT delayed recall\end{tabular} \\ \hline
Executive   function (working memory) & \begin{tabular}[c]{@{}l@{}}WAIS-III   \\ Digit Span Test – backward span, \\ Trail Making Test (Part B)\end{tabular} \\ \hline
Language   (semantic memory) & \begin{tabular}[c]{@{}l@{}}Multilingual Naming Test (MiNT), \\ Sematic fluency (animals), \\ Lexical/phonemic fluency\end{tabular} \\ \hline
Attention   (perceptual speed) & \begin{tabular}[c]{@{}l@{}}Trail Making Test (Part A), \\ WAIS-III   \\ Digit Span Test – forward span\end{tabular} \\ \hline
Visuospatial   ability & \begin{tabular}[c]{@{}l@{}}Clock-drawing test, \\ Copy of BVMT figures\end{tabular} \\ \hline
\end{tabular}
\caption{Testing battery for the five cognitive domains: episodic, working, semantic, perceptual and visuospatial.}
\label{tab: memory_tests}
\end{table}

\subsubsection{Outcomes and goal of the analysis:}
  The testing battery contains twelve performance tests related to five cognitive domains: episodic memory, executive function (working memory), language (semantic memory), attention (perceptual speed), visuospatial ability, recorded at baseline and each of the follow visits at the six months interval. The tests corresponding to the five cognitive domains are tabulated in Table~\ref{tab: memory_tests}.  Except for the trail making tests (Part B and Part A), higher score implies better cognition. Therefore, we multiply the trail making test scores by $-1$ to preserve `the higher the better' relationship between the outcomes and cognition. The raw scores of the twelve tests are centered around their baseline mean, and scaled by their baseline standard deviation, to obtain the standardized scores. The standardized test scores are further averaged within each domain to obtain the five-dimensional composite scores corresponding to the five cognitive domains, which serves as the response in our analysis. The actual date when the test scores are obtained for each subject is not exactly six months from the last time when the measurements are recorded. We consider the difference in days between the exact date of the recording of the test scores corresponding to $j$th visit from the baseline as $t_{ij}$ to model the composite test score as a smooth function over time for better understanding of cognitive decline. This implies $t_{i1}=0$ for all subjects in the study.

\subsection{Statistical analysis and findings}
In this section, we present the results and findings obtained by implementing the projection-based test on the TOMMORROW study.

\subsubsection{Implementation of projection-based test}
\begin{figure}
    \centering
    \includegraphics[scale=0.47]{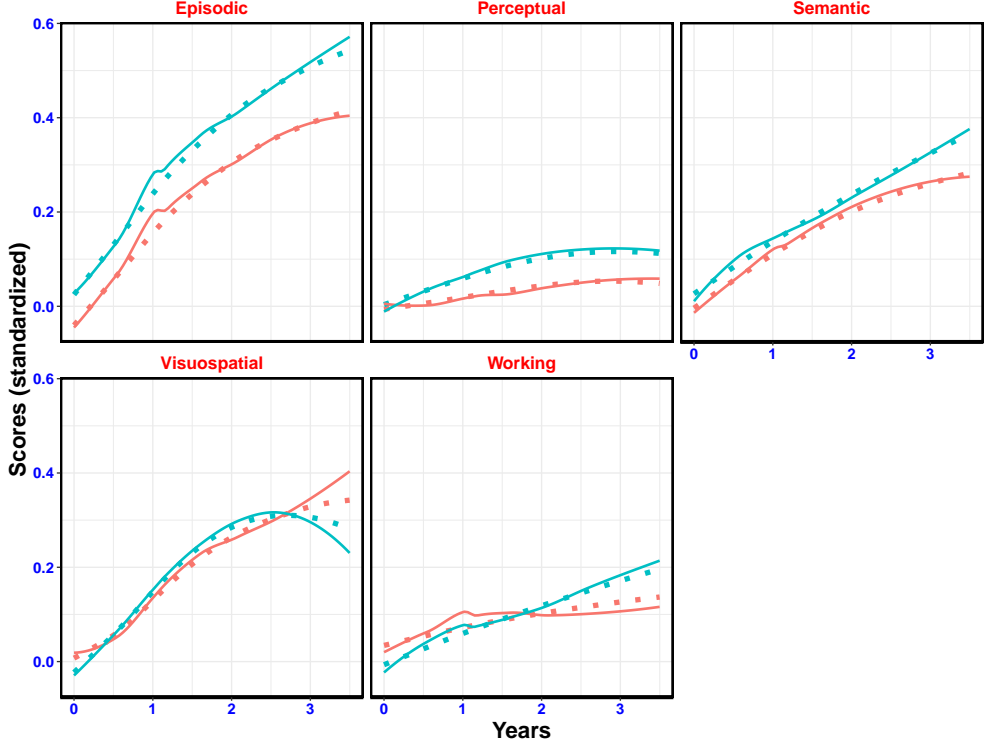}
\cprotect\caption{Locally weighted scatter plot smoothing (solid line) of the observed performance score, and the predicted smooth score (dotted line) for the pioglitazone (red) and the placebo (blue) group, across the five cognitive domain. Predicted scores are obtained {\verb|gam|()} function of {\verb|mgcv|} package in R.}
    \label{fig: Observed_vs_Predicted}
\end{figure}

Using five-dimensional standardized composite scores corresponding to five cognitive domain we fit  model~(\ref{eqn: model}) to implement our projection-based test. The mean part of the model is fitted by \verb|gam()| function in \verb|mgcv| package \citep{wood2011fast}, to estimate the overall smooth mean $\widehat{\mu}^{(\ell)}(t)$ as well as a smooth effect of pioglitazone $\widehat{\eta}^{(\ell)}(t)$. We do not include any baseline covariates in the model because the baseline analyses in Section \ref{sec: baselineanalysis} of the Supplementary material suggest that the two groups are homogeneous with respect to those baseline covariates. Figure \ref{fig: Observed_vs_Predicted} demonstrates the locally weighted average scatter plot (LOESS) smoothed version of the observed performance scores corresponding to the five cognitive domains, superimposed with the estimated mean response, i.e. $\widehat{\mu}^{(\ell)}(t)$ for the placebo group, and $\widehat{\mu}^{(\ell)}(t) + \widehat{\eta}^{(\ell)}(t)$, $\ell=1, \dots, q$ for the treatment group. The magnitude of the difference of the observed LOESS smoothed curve between the treatment and the placebo is small.


After estimating the mean components, we compute the residuals $\widetilde{\bE}_{ij} = \bY_{ij} - \widehat{\boldsymbol\mu}(t_{ij}) - g_i\widehat{\boldsymbol\eta}(t_{ij})$ for each $i, j$. Using the residuals we estimate the smooth covariance function  using R package \verb|mfaces|  \citep{mfaces}, and the leading eigenfunctions $\{\widehat{\bpsi}_k(t)\}_k$ by spectral decomposition of the estimated covariance. We obtain $K = 8$ principal directions by setting a PVE equal to $99\%$. Corresponding to the eight estimated directions, we obtain the $8$-dimensional BLUP estimated mv-fPC scores $\widehat{\bxi}_i = (\widehat{\xi}_{i,1}, \dots, \widehat{\xi}_{i,8})^\top$ for each participant, $i=1,\dots, 1947$. Finally we implement a multivariate Hotelling $T$-squared test using the scores between the treatment and placebo group under the assumption of equal variance (equation~(\ref{eqn: HotellingTestRule_eqvar})) and unequal variance (equation~(\ref{eqn: HotellingTestRule_uneqvar})). Table~\ref{tab: pvalue_hotelling} presents the test-statistic and p-value of the test based on both the F-quantile as well as based on permutation test All the p-values are similar, and they suggest that there is no significant difference between cognitive scores the pioglitazone and the placebo group across all the cognitive domains. This corroborates to the finding obtained by the other results published based on the TOMMORROW study, that pioglitazone did not convey any significant benefit to the cognitive performance as compared to placebo group \citep{burns2021safety}.

\begin{table}[h]
\centering
\begin{tabular}{lccc}
\hline
\multicolumn{1}{c}{\multirow{2}{*}{Variance}} & \multirow{2}{*}{${T}_n$} & \multicolumn{2}{c}{p-value} \\ \cline{3-4} 
\multicolumn{1}{c}{} &  & F-test & Permutation \\ \hline
Equal & 10.64 & 0.226 & 0.223 \\
Unequal & 10.65 & 0.226 & 0.226 \\ \hline
\end{tabular}
\caption{Test-statistic, and the p-value of the Hotelling $T$-squared test for both the F-based cutoff, and the permutation based test. The tests were conducted in two setup: i) variance are equal ii) variance unequal. P-value of the permutation test is obtained based on $10000$ samples from the null distribution.}
\label{tab: pvalue_hotelling}
\end{table}

\begin{figure}
    \centering
    \includegraphics[scale=0.47]{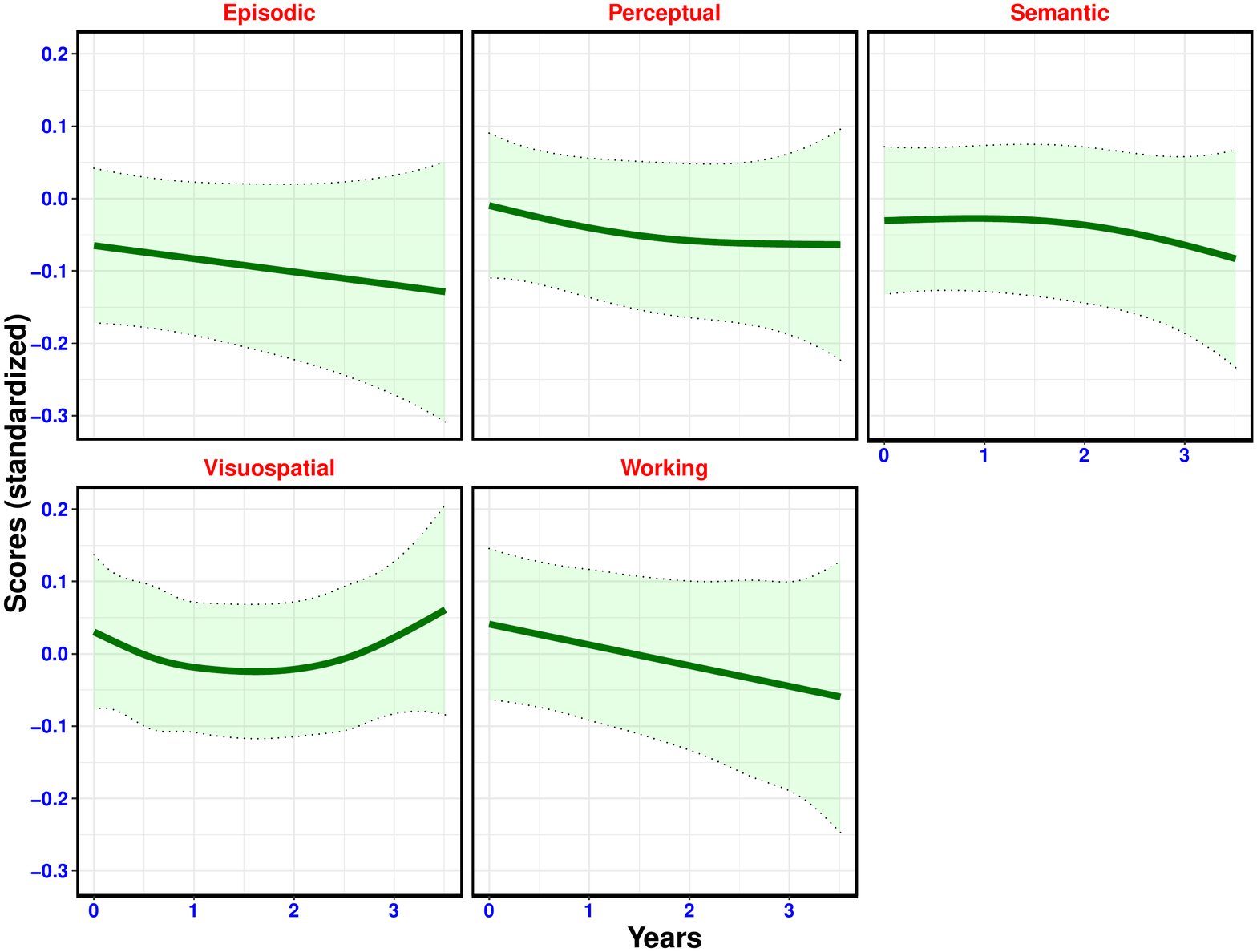}
    \caption{$95\%$ bootstrap confidence band (shaded in faded green) for the difference between the mean performance score of pioglitazone and the placebo group, for the five cognitive domain. The thick line is the estimated treatment effect $\widehat{\eta}^{(\ell)}(t)$.}
    \label{fig: EstTreatment_with_CB}
\end{figure}

Figure~\ref{fig: EstTreatment_with_CB} presents the estimated smooth effect of the treatment for all the five cognitive domains along with the $95\%$ bootstrap standard error-based simultaneous confidence band (shaded in light green). The results are presented based on $10,000$ bootstrap sample of subjects. As all of the confidence bands contain the zero function, it indicates that there is no significant pioglitazone effect. This further strengthens the conclusion drawn from the projection-based test. 


\section{Discussion}  \label{sec: mface_testing_discussion}
In this paper, we present a projection-based two-sample inference procedure for sparsely observed multivariate functional data. The main advantage of this test is that it returns a single p-value to determine whether there is any significant difference in the mean function in any of the $q$-components. Thus, it overcomes the issue of adjusting for multiple p-values that arises due to comparing the means in each of components separately. We are hopeful that our projection-based testing methodology will be increasingly applied to longitudinally designed randomized clinical trials where multiple secondary endpoints are collected to test for the treatment efficacy between two or more arms, as well as observational studies.   

Our testing procedure reduces the dimension of the infinite-dimensional response by an application of multivariate fPCA, which is popular in fDA literature. The major computational complexity is driven by the calculation involved in conducting fPCA for multivariate data, which grows at a cubic rate with dimension of the response $q$. To this end, throughout the article we have assumed that the dimension of the multivariate response is fixed. However, when the dimension of the response gets arbitrarily large, new computational techniques will be necessary to conduct the fPCA. Extension this procedure to the case when $q$ grows to $\infty$ requires further attention, as one might need to impose further sparsity condition to carry out the multivariate fPCA. Once the fPCA is carried out, computation of the test-statistic and the p-value is fast.

The proposed testing procedure provides a fresh outlook for two-sample inference problem in longitudinal design when the responses are multidimensional. Although for a sparse data, the scores are estimated under a Gaussian assumptions, our simulation study shows that the test performs quite well for non-Gaussian data. It is important to note that the performance of the testing procedure hinges upon the quality of the estimation of the eigencomponents. Therefore, rate of convergence of the test-statistic to the asymptotic null distribution will primarily depend on the rate at which the covariance components are estimated under a sparse design. 

While representing the response trajectories through multivariate fPCA representation, we have inherently assumed that all the outcomes are measured in the same units, and have similar range of variation. In Section~\ref{sec: mface_testing_azillectstudy} we have standardized the outcomes to ensure this. However, if a particular component has relatively large variability, we can represent the functions using normalized multivariate fPCA developed by \citep{chiou2014multivariate}, and conduct the rest of the testing procedure using the normalized fPC scores. 

Another advantage of our projection-based testing procedure is that it can be extended in principle to the case of mv-fD with more complex structure, such as when the components functions are observed over heterogeneous domain (such as data consisting of unidimensional functions, and images) by an application of the mv-fPCA method developed by \cite{happ2018multivariate}. Similarly, one can apply the testing procedure to the two-sample inference problem of manifold valued multivariate functional data, by application of fPCA on Riemannian manifolds \citep{dai2018principal}. 

\section*{Software}
\label{sec: software}

Software in the form of R code for the entire simulation study is available publicly on 
\url{https://github.com/SalilKoner/ProjectionTesting}.

\section*{Funding}
Support for this work comes from National Institute on Aging (grants AG064803, P30AG072958, and P30AG028716 to Sheng Luo) and from the National Philanthropic Trust- Bill and Melinda Gates Foundation/Alzheimer’s Drug Discovery Initiative (ADDI)

\section*{Acknowledgements}
This data used in this study was from the TOMMORROW trial 
(\url{https://clinicaltrials.gov/ct2/show/NCT02932410}). The authors would like to gratefully acknowledge the volunteer participants, the project partners, and the clinical site investigators and staff of the TOMMORROW trial. The TOMMORROW Trial (NCT01931566) was sponsored by Takeda Pharmaceutical Company in collaboration with Zinfandel Pharmaceutical Company (Chapel Hill, NC).  The authors would also like to acknowledge the project team at Duke Clinical Research Institute (DCRI), specifically Rebecca Wilgus MS, Dr. Jack Shostak, and Dr. Weibing Xing for their assistance in accessing  and navigating the TOMMORROW datasets used in these analyses.

\bibliographystyle{biorefs}
\bibliography{refs_main}
 
\section*{Supplementary Material}
\label{sec: supp}

\renewcommand{\theequation}{S\arabic{equation}} 
\renewcommand{\thesection}{S\arabic{section}}  

The supplementary material contains a detailed description for the estimation of the multivariate functional principal scores (mv-fPC) that serves as the building block for our projection-based test, as well as additional supporting documents for the TOMORROW study analysis.

\subsection{Estimation of data-driven projections}
\label{sec: mface_estimation_of_eigenfun}

For sparsely observed functional data, the scores are obtained via best linear unbiased prediction (BLUP) under the mixed model \citep{yao2005functional}
\begin{equation} \label{eqn: mixedmodel}
    \bY_i = {\boldsymbol{\mu}}_i + \sum_{k=1}^{{K}}
 {\boldsymbol{\psi}}_{i,k} {\xi}_{i,k} +\boldsymbol{\epsilon}_i,
\end{equation}
where $\bY_i = (Y^{(1)}_{i1}, \dots, Y^{(1)}_{im_i}, \dots, Y^{(q)}_{i1}, \dots, Y^{(q)}_{im_i})^\top$ be the $q \times m_i$-length stacked vector of response for the $i$th subject, ${\boldsymbol{\mu}}_{i,k} = ({\boldsymbol{\mu}}^{(1)^\top}_{i,k}, \dots, {\boldsymbol{\mu}}^{(q)^\top}_{i,k})^{\top}$ with ${\boldsymbol{\mu}}^{(\ell)}_{i,k} = (\mu^{(\ell)}(t_{i1}), \dots, \mu^{(\ell)}(t_{im_i}))^\top$ and ${\boldsymbol{\psi}}_{i,k} = ({\boldsymbol{\psi}}^{(1)^\top}_{i,k}, \dots, {\boldsymbol{\psi}}^{(q)^\top}_{i,k})^{\top}$ with ${\boldsymbol{\psi}}^{(\ell)}_{i,k} = (\psi^{(\ell)}(t_{i1}), \dots, \psi^{(\ell)}(t_{im_i}))^\top$ are the $q \times m_i$-length stacked vector of mean and eigenfunctions evaluated at the observations points of the $i$th subject, and $\boldsymbol\epsilon_i$ are the measurement error. Under the Gaussian assumption of the scores and the measurement error, the best linear unbiased predictor (BLUP) of $\boldsymbol\xi_i$ under the model~(\ref{eqn: mixedmodel}) is of the form
\begin{equation} \label{eqn: BLUP}
    \widetilde{\boldsymbol\xi}_i := \mathbb{E}(\boldsymbol\xi_i \mid \bY_i) = \text{diag}(\lambda_1, \dots, \lambda_K) {\boldsymbol\Psi}_i^{\top} \bG_{\bY_i}^{-1} (\bY_i - \boldsymbol\mu_i),
\end{equation}
where $\bG_{\bY_i} = \text{Cov}(\bY_i) = \{\Sigma_{\ell, \ell^\prime}(t_{ij}, t_{ij^\prime}) + \tau_\ell^2 \mathbb{I}(j = j^\prime,\; \ell= \ell^\prime)\}_{1 \leq \ell, \ell^\prime\leq q,\; 1 \leq j, j^\prime \leq m_i}$ be the covariance matrix of $\bY_i$ and $\boldsymbol\Psi_i =(\bpsi_{i,1}, \dots, \bpsi_{i,K})$ be the column-stacked version of $\{\bpsi_{i,k}\}_{k=1}^K$'s. The BLUP estimator $\widetilde{\boldsymbol\xi}_i$ is a consistent estimator of the true scores $\boldsymbol\xi_i$ as the number of observations per subject grows and the measurement error gets small. It is also important to note that unlike the true scores, $\{\widetilde{\xi}_{i,k}\}_{k=1}^K$ are also not uncorrelated across $k$. 

The BLUP estimator in equation~(\ref{eqn: BLUP}) is empirically obtained by plugging in the estimator of $\bmu_i$, $\bG_{\bY_i}$, $\{\lambda_k\}_{k=1}^K$ and $\boldsymbol\Psi_i$. This requires estimation of the covariance function $\{\Sigma_{\ell, \ell^\prime}(t,t^\prime)\}_{1 \leq \ell, \ell^\prime \leq q}$, and the measurement error variance $\{\tau_\ell^2\}_{\ell=1}^q$ as well as the eigencomponents of the covariance. First, a smooth estimator of the mean function $\mu^{(\ell)}(t)$ is obtained by smoothing the response of $\ell$th coordinate $\{Y_{ij}^{(\ell)} : j=1,\dots, m_i\}_{i=1}^n$ for all the subjects under working independence assumption to obtain the residual $E_{ij}^{(\ell)} = Y_{ij}^{(\ell)} - \widehat{\mu}^{(\ell)}(t_{ij})$. For the sparse data, one needs to pool the data from all the subjects to obtain a smooth estimator of $\bSigma(t,t^\prime)$. Unlike univariate fD, this requires estimation of large covariance matrix. A computationally efficient implementation for the estimation of $\bSigma(t,t^\prime)$ is given in \cite{li2020fast}. Briefly, the cross-covariance function is modeled as a tensor product of univariate bases, $\Sigma_{\ell\ell^\prime}(t,t^\prime) = \mathbf{B}^\top(t)\boldsymbol\Gamma_{\ell\ell^\prime}\mathbf{B}(t^\prime)$, where $\mathbf{B}(t) = (B_1(t), \cdots, B_r(t))^{\top}$ is the $r$-length vector of basis functions and $\boldsymbol\Gamma_{\ell\ell^\prime}$ is a matrix of coefficients; A penalized regression estimator of $\boldsymbol\Gamma_{\ell\ell^\prime}$ is obtained by the ``raw covariances" $\{\widetilde{E}^{\ell}_{ij}\widetilde{E}^{\ell^\prime}_{ij^\prime}\}_{i,j}$ as pseudo responses, for all $1 \leq \ell \leq \ell^\prime \leq q$ and setting  $\boldsymbol\Gamma_{\ell\ell^\prime}=\boldsymbol\Gamma_{\ell^\prime\ell}^\top$. The eigenfunctions are efficiently estimated as $\widehat{\boldsymbol\psi}_k (t) = (\mathbf{I}_q \otimes \mathbf{B}^\top(t) \mathbf{S}^{-\frac{1}{2}} ) \widehat{\mathbf{V}}_k$, where $\mathbf{S}=\int \mathbf{B}(u) \mathbf{B}^\top(u)du$ and $\{\widehat{\mathbf{V}}_k\}_k$ are eigenvectors of the block matrix with elements $\mathbf{S}^{\frac{1}{2}} \widehat{\boldsymbol\Gamma}_{\ell\ell^\prime}   \mathbf{S}^{\frac{1}{2}}$, for $1\leq \ell, \ell'\leq q$. The optimal number of eigenfunctions $\widehat{K} =  \widehat{K}(n)$ are chosen via as the minimum value of $K$ so that percentage of variation of explained (PVE), defined as $\sum_{k=1}^K\widehat{\lambda}_{k}/\sum_{k=1}^\infty\widehat{\lambda}_{k}$, is higher than some pre-specified threshold. Other methods for optimally choosing $K$ include cross-validation, Akaike information criterion (AIC) and Bayesian information criterion (BIC). 

Once the eigencomponents $\{\widehat{\lambda}_k, \widehat{\bpsi}_k(t)\}_{k=1}^{\widehat{K}}$ are estimated from the data, the scores are empirically estimated by plugging in the estimator of the mean and the variance components in equation~(\ref{eqn: BLUP}), to obtain
\begin{equation} \label{eqn: BLUP_emp}
    \widehat{\boldsymbol\xi}_i =  \widehat{\mathbf{V}}_{k}^{\top}(\mathbf{I}_q \otimes \mathbf{S}^{\frac{1}{2}} )\widehat{\boldsymbol\Gamma}(\mathbf{I}_q \otimes \bB_i^\top) \widehat{\bG}_{\bY_i}^{-1} (\bY_i - \widehat{\boldsymbol\mu}_i),
\end{equation}
where $\bB_i = (\bB(t_{i1}), \cdots, \bB(t_{im_i}))^\top$, $\widehat{\boldsymbol\Gamma} = ((\widehat{\boldsymbol\Gamma}_{\ell, \ell^\prime}))_{1 \leq \ell, \ell^\prime \leq q} \in \Real^{qr \times qr}$ and $\widehat{\bG}_{\bY_i} = (\mathbf{I}_q \otimes \bB_i)\widehat{\boldsymbol\Gamma}(\mathbf{I}_q \otimes \bB_i^\top) + \textrm{blockdiag}(\widehat{\tau}^2_{1}\mathbf{I}_{m_{i}}, \dots, \widehat{\tau}^2_{q}\mathbf{I}_{m_{i}})$. Thus, this method bypasses the inversion of large $q \times m_i$ by  $q \times m_i$ covariance matrix for each $i=1, \dots, n$ by inverting the $qr \times qr$ matrix $\widehat{\bGamma}$ only once, reducing the computational burden involved in score estimation of mv-fD, especially when $q$ is relatively large.

\subsection{Additional results for supporting the TOMORROW study analysis} \label{sec: baselineanalysis}
In this section, we will present some additional tables related to the baseline analysis of TOMORROW study.
\begin{table}
\centering
\resizebox{0.7\textwidth}{!}{%
\begin{tabular}{|lccc|}
\hline
\cellcolor[HTML]{A7EFF1} & \cellcolor[HTML]{A7EFF1} & \cellcolor[HTML]{A7EFF1} & \cellcolor[HTML]{A7EFF1} \\
\cellcolor[HTML]{A7EFF1} & \cellcolor[HTML]{A7EFF1} & \cellcolor[HTML]{A7EFF1} & \cellcolor[HTML]{A7EFF1} \\
\multirow{-3}{*}{\cellcolor[HTML]{A7EFF1}} & \multirow{-3}{*}{\cellcolor[HTML]{A7EFF1}\textbf{\begin{tabular}[c]{@{}c@{}}High risk \\ pioglitazone \\ (N = 985)\end{tabular}}} & \multirow{-3}{*}{\cellcolor[HTML]{A7EFF1}\textbf{\begin{tabular}[c]{@{}c@{}}High risk \\ placebo\\ (N = 962)\end{tabular}}} & \multirow{-3}{*}{\cellcolor[HTML]{A7EFF1}\textbf{\begin{tabular}[c]{@{}c@{}}Overall\\ (N = 1947)\end{tabular}}} \\ \hline
\multicolumn{4}{|l|}{\cellcolor[HTML]{FFCCC9}\textbf{Age (in years)}} \\
Mean (SD) & 74.7 (5.20) & 74.9 (5.19) & 74.8 (5.19) \\
Range & 65 - 83 & 65 - 83 & 65 - 83 \\
\multicolumn{4}{|l|}{\cellcolor[HTML]{FFCCC9}\textbf{Sex}} \\
Female & 543 (55.1\%) & 579 (60.2\%) & 1122 (57.6\%) \\
Male & 442 (44.9\%) & 383 (39.8\%) & 825 (42.4\%) \\
\multicolumn{4}{|l|}{\cellcolor[HTML]{FFCCC9}\textbf{Body mass index (in mg/m2)}} \\
Mean (SD) & 28.0 (5.15) & 28.0 (5.35) & 28.0 (5.25) \\
Range & 14 - 68 & 16 - 52 & 14 - 68 \\
Missing & 2 (0.2\%) & 1 (0.1\%) & 3 (0.2\%) \\ \hline
\end{tabular}%
}
\caption{Demographic characteristics of the participants at baseline for the two arms in the TOMORROW study.}
\label{tab: baselineanalysis}
\end{table}

\subsubsection*{Baseline analysis:} 
The average age of the participants enrolled in the study is $74$ years with a standard deviation of $5$ years, which is similar to both the pioglitazone and the placebo group. About $60\%$ of the individuals in the cohort are over $75$ years of age in both the groups. About $55\%$ of the population in pioglitazone group are women, whereas about $60\%$ are women in the placebo group. The baseline height (in cm), weight (kg) and the body mass index (BMI) (in mg/m2) are within similar range across the groups with mean height about $167$ cm (SD $10$ cm), mean weight of $78$ kgs (SD $17$ kgs) and mean BMI of $28$ mg/m2 (SD $5$ mg/m2) respectively. The baseline analyses tabulated in Table~\ref{tab: baselineanalysis} suggest that the two groups are comparable with respect to key baseline demographics.

\end{document}